\documentclass[sigconf]{acmart}
\usepackage{amsmath}
\usepackage{pifont}
\AtBeginDocument{%
  }

\copyrightyear{2025}
\acmYear{2025}
\setcopyright{rightsretained}
\acmConference[RecSys '25]{Proceedings of the Nineteenth ACM Conference on Recommender Systems}{September 22--26, 2025}{Prague, Czech Republic}
\acmBooktitle{Proceedings of the Nineteenth ACM Conference on Recommender Systems (RecSys '25), September 22--26, 2025, Prague, Czech Republic}\acmDOI{10.1145/3705328.3748096}
\acmISBN{979-8-4007-1364-4/2025/09}

\title{USD: A User-Intent-Driven Sampling and Dual-Debiasing Framework for Large-Scale Homepage Recommendations}

\author{Jiaqi Zheng}
\author{Cheng Guo}
\email{zhengjiaqi.zjq@taobao.com}
\email{mike.gc@taobao.com}
\affiliation{%
  \institution{Taobao \& Tmall Group of Alibaba}
  \city{Beijing}
  \country{China}
}

\author{Yi Cao}
\authornote{Corresponding author.}
\author{Chaoqun Hou}
\author{Tong Liu}
\email{dylan.cy@taobao.com}
\email{hcq.hcq@taobao.com}
\email{yingmu@taobao.com}
\affiliation{%
  \institution{Taobao \& Tmall Group of Alibaba}
  \city{Hangzhou}
  \country{China}
}

\author{Bo Zheng}
\email{bozheng@alibaba-inc.com}
\affiliation{%
  \institution{Taobao \& Tmall Group of Alibaba}
  \city{Beijing}
  \country{China}
}

\begin{abstract}
Large-scale homepage recommendations face critical challenges from pseudo-negative samples caused by exposure bias, where non-clicks may indicate inattention rather than disinterest. Existing work lacks thorough analysis of invalid exposures and typically addresses isolated aspects (e.g., sampling strategies), overlooking the critical impact of pseudo-positive samples - such as homepage clicks merely to visit marketing portals. We propose a unified framework for large-scale homepage recommendation sampling and debiasing. Our framework consists of two key components: (1) a user intent-aware negative sampling module to filter invalid exposure samples, and (2) an intent-driven dual-debiasing module that jointly corrects exposure bias and click bias. Extensive online experiments on Taobao demonstrate the efficacy of our framework, achieving significant improvements in user click-through rates (UCTR) by 35.4\% and 14.5\% in two variants of the marketing block on the Taobao homepage, Baiyibutie and Taobaomiaosha.

\end{abstract}


\begin{document}
\maketitle

\section{Introduction}
\label{intro}

\begin{figure*}[h]
  \centering
  \includegraphics[width=0.9\linewidth,height = 0.308\linewidth]{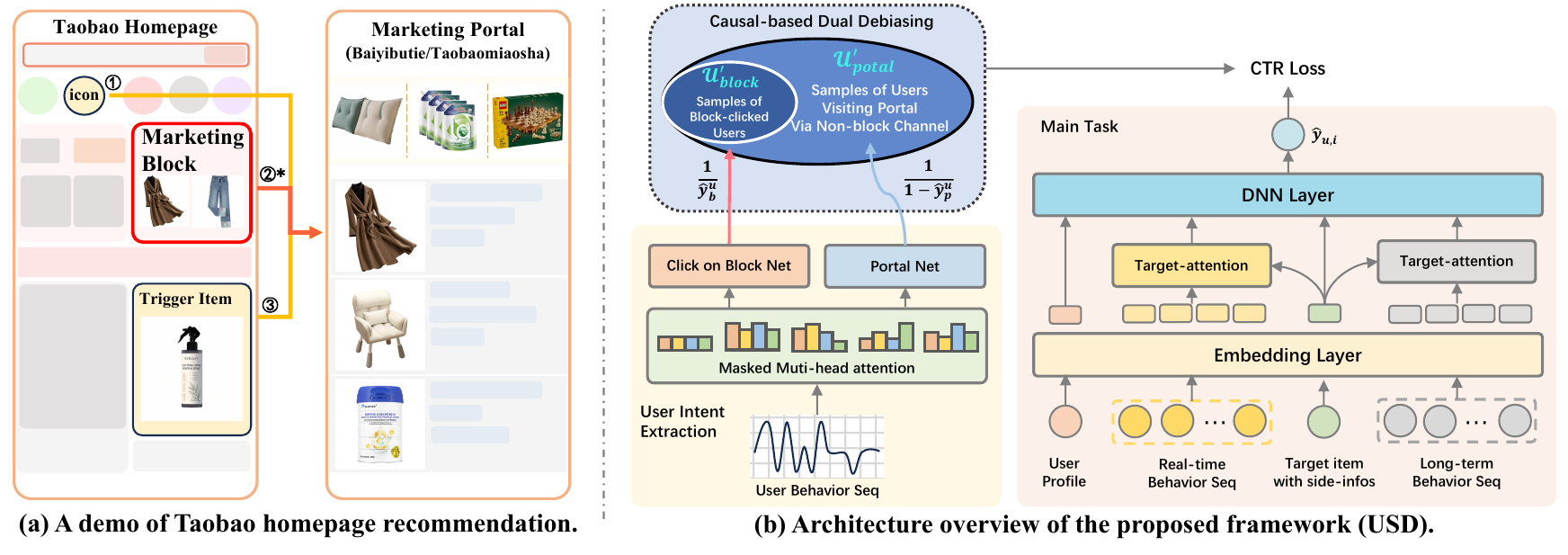}
  \caption{Model architecture of USD and Taobao homepage recommendation demonstration. (a) Clicking any of the three entry points (\ding{173}* denotes the marketing block) in the homepage leads to the marketing portal. (b) USD structure with User Intent Extraction and Causal-Based Dual-Debiasing modules.}
  \Description{Architecture overview of USD.}
  \label{modelpic}
\end{figure*}

Modern e-commerce recommendation systems (RS) employ homepage interfaces to serve hundreds of millions of daily active users. 
Homepage blocks, serving as one of the key entry points for user traffic redirection, exhibit two critical characteristics: 
1. Clicks redirect users to a specific section (e.g., marketing portal), as shown in Fig.~\ref{modelpic}(a). 
%2. Diverse user behavior patterns with attention constraints lead to inevitable item overlooking.
2. Diverse user behaviors under attention constraints, particularly goal-driven interactions like product searches, inherently cause exposed item neglect.
These create massive invalid exposures and biased samples misrepresenting true user interests, introducing substantial noise into the RS.
Current research lacks joint mitigation of invalid exposure and sample selection bias (SSB) in homepage scenarios.
Sample reliability studies predominantly follow two paradigms: sampling-based and debiasing methods.
% In contemporary e-commerce ecosystems, homepage product presentation constitutes a important component of personalized recommendation systems, serving hundreds of millions of daily active users (DAUs). 
% However, the inherent diversity of user behavior patterns coupled with limited attention spans creates a significant challenge: users inevitably overlook a substantial portion of recommended products, leading to the generation of massive false exposure samples within the daily billions of homepage impressions. These false exposures, which fail to accurately reflect genuine user interests, introduce considerable noise into the recommendation system. 
% Despite significant advancements in recommendation systems, a critical research gap persists in jointly addressing pseudo exposure and sample confidence bias within large-scale homepage scenarios. Research on sample confidence have predominantly bifurcated into two methodological paradigms: sample sampling-based methods and debiasing-based methods.

\textbf{Sampling-based methods} primarily use statistical heuristics: fairness-optimized~\cite{DBLP:conf/www/ChenFCLLZL23, 10.1145/3459637.3481948}, distribution-based~\cite{DBLP:conf/nips/DingQY0J20, 10.1145/3539597.3570419}, and in-/cross-batch strategies~\cite{Yang2020MixedNS,DBLP:journals/corr/HidasiKBT15,DBLP:conf/cikm/HidasiK18,10.1145/3366424.3386195, Wang2021CrossBatchNS}. 
While most existing works enhance sample quality through pseudo-negative mining~\cite{DBLP:conf/sdm/LiuLWZZLCF23,10.1145/2872427.2883090}, they ignore user intent.
Few works incorporate behavioral intentions~\cite{10.1145/3604915.3608791,chen2024intentenhanceddataaugmentationsequential} but focus on representation learning and sample generation rather than direct sampling guidance.
% Existing approaches improve sample quality via pseudo-negative mining~\cite{DBLP:conf/sdm/LiuLWZZLCF23,10.1145/2872427.2883090} but neglect user intent.
% Behavior-aware methods like~\cite{10.1145/3604915.3608791} use historical interactions for pseudo-negative filtering but focus on representation learning over sampling.
Effective exposure bias mitigation requires user-centric interaction analysis to identify authentic attention, combined with intent-explicit joint debiasing strategies.

\textbf{Debiasing methods} address specific biases (e.g., selection and exposure bias)~\cite{DBLP:journals/tois/0007D0F0023}.
Existing approaches enhance sampling via importance or popularity weighting~\cite{DBLP:conf/kdd/LianWG0C20,DBLP:conf/www/LiangCMB16} or counterfactual space/pseudo-labels~\cite{DBLP:conf/icde/ZhuZYL0ZZ0W0W23,10.1145/3640457.3688151} for SSB mitigation, yet ignore user guidance.
Recent advances adopt user-aware strategies~\cite{chen2020fast,DBLP:conf/www/ZhengGLHLJ21,10.1145/3308558.3313582,Wang2020SamWalkerRW}.
ESMM~\cite{10.1145/3209978.3210104} and ESCM$^2$~\cite{10.1145/3477495.3531972} use behavioral patterns and inverse propensity weighting for debiasing but neglect intention diversity.
We argue personalized debiasing requires individual intentions from behavioral patterns for unified bias mitigation and intent modeling.

To address these challenges, we propose a \textbf{U}ser-Intent-Driven \textbf{S}ampling and \textbf{D}ual-Debiasing framework (USD) for large-scale homepage recommendations. 
% We extract confidence samples through authentic user interactions and leverage users' historical behavior sequences to model personalized intents and further perform dual-debiasing.
The main contributions of this work are summarized as follows:
\begin{itemize}
\item To our knowledge, this is the first work jointly addressing invalid exposures and SSB in large-scale homepage recommendations through intent-driven sampling and debiasing, delivering a scalable-effective production solution.
\item We propose an intent-aware sampler to select confident samples from large-scale datasets under severe invalid exposure, along with a dual-debiasing module for fine-grained debiasing via differentiated rectification.
\item Extensive offline experiments on Taobao's homepage data demonstrate USD's state-of-the-art performance.
Online A/B tests show 35.4\% and 14.5\% UCTR gains for Taobao's Baiyibutie and Taobaomiaosha homepage blocks, respectively.
\end{itemize}
\section{Methods}

\subsection{Problem Definition}

% We define the entire exposed user set as $\mathcal{U}=\{u_1,u_2,\ldots,u_n\}$ and the set of all exposed items as $\mathcal{I}=\left\{\textit{i}_1,\textit{i}_2,\ldots,\textit{i}_m\right\}$, where $n$ and $m$ denote the number of exposed users and items respectively. 
We formulate the task as a click-through rate (CTR) prediction model, where $\mathcal{U}$ and $\mathcal{I}$ denote exposed users and items. $\mathcal{Y} = \mathcal{U} \times \mathcal{I}$ represents user-item click labels, and $\hat{y}_{u,i} \in [0,1]$ is the predicted probability for user $u$. 
Sampling is modeled as $\mathcal{U}'=F_s( \mathcal{U} )$, where $\mathcal{U}'$ is the confidence subset of $\mathcal{U}$ and $F_s$ is the sampling function. To learn fine-grained intent distributions $\hat{y}_p^u$ and $\hat{y}_b^u$, we introduce an auxiliary debiasing framework $F_{\theta}$, supervised by $\mathcal{Y}_p$ and $\mathcal{Y}_b$. Here, $y_p^u,y_{b}^{u} \in \{0,1\}$ indicates whether user $u$ visited the marketing portal or clicked the marketing block on the current day.

% \subsection{Model Overview}

% To address the challenge of handling hundreds of millions of pseudo-exposed samples, 
% we propose a unified framework integrating joint sampling with fine-grained debiasing based on user intents.
% As illustrated in Fig.~\ref{modelpic}(b), our framework focuses on two principal objectives: (1) the CTR main task and (2) two auxiliary tasks designed to learn user intents toward the marketing portal and marketing blocks. 
% The model consists of two key components. The first component implements intent-aware confidence sampling. Specifically, we filter out users with no relevant intent by analyzing their behavioral patterns across all exposed users, ensuring the reliability of the sampled training and evaluation data.
% Given marketing blocks' role as homepage traffic entry points and users' divergent intents toward blocks vs portals (e.g., clicking block merely to visit the portal), we further decouple intent modeling for these two entities. Additionally, we apply differentiated dual-debiasing on the sampled data via Inverse Propensity Scoring (IPS)~\cite{10.5555/3045390.3045567}. The details of this approach will be elaborated in subsequent subsections.

\subsection{User Intent-Driven Negative Sampling} \label{subsc:sampling}
Dynamic Negative Sampling has interpretability and stability issues~\cite{ma2024negative}, while random sampling on large-scale exposure data causes significant loss of confident negatives. Thus, we leverage real user interactions for direct sampling guidance. 

As mentioned in Sec.~\ref{intro}, item clicks lead users to the marketing portal.
Prolonged exposure to this interaction pattern establishes entrenched cognitive schemas, aligning users' intents toward the marketing block with those toward the portal.
Temporal dynamics in user intent patterns further show that users' same-day portal engagements correlate more strongly with user intent than cross-day counterparts. We thus propose the following sampling mechanism: 
\begin{equation}
    \mathcal{U}'=\{u\in\mathcal{U} \mid y_p^u =1 \},
\end{equation}
where $y_p^u=1$ indicates user $u$ visited the marketing portal on the current day. Our sampling module extends click-based sampling by incorporating multi-channel entry points (e.g., icons and notification triggers) that capture consistent latent user intents toward marketing portal.
This approach outperforms click-based sampling methods in reliability, as validated in Sec.~\ref{ablationstudy}.

\subsection{User Intent Extraction Module}
\label{UIEM}
The marketing block acts as an entry point for homepage traffic, with user clicks reflecting either item interest or intent to visit the marketing portal.
Therefore, we construct the User Intent Extraction Module (UIEM) to model users' fine-grained intent. 
Empirical analysis reveals that user intents exhibit temporal dynamics with periodicity, making historical behavior sequences better capture user intents compared to static user data.
Let $\mathcal{S}$ denote users' past-month behavioral sequences, where $s_u^{t}\in\{-1,0,1\}$: $-1=$ no portal visit, $0=$ non-block portal visit, $1=$ block click.
% Specifically, $s_u^{t}=-1$ indicates that user $u$ did not visit the marketing portal on day $t$, $s_u^{t}=0$ signifies that user $u$ visited the marketing portal on day $t$ via non-homepage block channels, and $s_u^t=1$ indicates that user $u$ clicked marketing block on day $t$. 
We then define user's intent probabilities for the portal and block as follows:
\begin{align}
\hat{y}_p^u &= \sigma(\mathrm{MLP}_\text{portal}(H)), \\
\hat{y}_b^u &= \sigma(\mathrm{MLP}_\text{block}(H)),    
\end{align}
where $\sigma(\cdot)$ is the activation function, $\mathrm{MLP}(\cdot)$ denotes a multi-layer perceptron, and $H$ is the user behavioral features computed as:
\begin{equation}
    H= \mathrm{Decoder}(E_s+E_\mathrm{pos})+\mathrm{Pooling}(E_s).
\end{equation}
Here, $E_s$ is the embedding of the user's behavioral sequence. The $\text{Decoder}(\cdot)$ incorporates the causal mask and positional encoding $E_\mathrm{pos}$, following the Transformer architecture~\cite{10.5555/3295222.3295349}. 
Notably, $\hat{y}_b^u$ denotes the probability of a user interacting with the entire marketing block, including its title and two displayed items.

To address invalid exposures and sampling-scale considerations, we propose a strategic sampling approach by selecting users who visited the marketing portal in the past week as UIEM training samples. 
The sampling process is defined as: $\hat{\mathcal{U}}=\{u\in\mathcal{U} \mid r_p^u =1 \}$, where $r_p^u=1$ indicates user $u$ visited the marketing portal in the past week. Binary labels $y_{p}^{u}$ and $y_{b}^{u}$ supervise the auxiliary task. UIEM's loss functions are formulated as follows:
\begin{align}
\mathcal{L}_\mathrm{portal} &= \frac{1}{|\hat{\mathcal{U}}|} \sum_{u\in\hat{\mathcal{U}}} \left[-y_{p}^{u}\log\hat{y}_p^u-(1-y_p^{u})\log(1-\hat{y}_p^u)\right],\\
\mathcal{L}_\mathrm{block} &= \frac{1}{| \hat{\mathcal{U}} |} \sum_{u\in\hat{\mathcal{U}}} \left[ -y_b^{u}\log\hat{y}_b^u-(1-y_b^{u})\log(1-\hat{y}_b^u) \right].
\end{align}

\subsection{Causal-Based Dual-Debiasing Module}
\label{CDM}
The sampled set $\mathcal{U}'$ remains biased due to two factors.
First, marketing blocks' traffic entry characteristics induce click bias through interactions potentially generated solely for portal visits.  
Second, users with strong marketing portal intent but no clicks indicate the recommended items fail to align with their interests - these critical negative samples deserve higher weights than others. 
We partition $\mathcal{U}'$ into $\mathcal{U}_{\mathrm{portal}}'$ (users visiting the marketing portal via non-homepage-block channels) and $\mathcal{U}_{\mathrm{block}}'$ (marketing block click users).
Subsequently, we perform the inverse propensity scoring (IPS)~\cite{10.5555/3045390.3045567} based dual-debiasing method using the outputs $\hat{y}_p^u$ and $\hat{y}_b^u$ from the UIEM (Sec.~\ref{UIEM}). The debiased CTR loss is defined as:
\begin{equation}
{\mathcal{L}}_\text{CTR}' =\begin{cases} 
\begin{aligned}
 &\frac{1}{|\mathcal{U}'|}\sum\limits_{(u,i)\in\mathcal{U}'} 
 \frac{e(y_{u,i},\hat{y}_{u,i})}{1-\hat{y}_p^u}, u \in \mathcal{U}'_\text{portal}\\
 & \frac{1}{|\mathcal{U}'|}\sum\limits_{(u,i)\in\mathcal{U}'}
 \frac{e(y_{u,i},\hat{y}_{u,i})}{\hat{y}_b^u}, u \in \mathcal{U}'_\text{block}
\end{aligned}\end{cases}
\label{eq:loss-ctr},
\end{equation}
where $\hat{y}_{u,i}$ denotes the click-through rate predicted by our baseline model \textbf{BASE}, whose architecture is analogous to ETA~\cite{Chen2021EndtoEndUB}. The cross-entropy loss $e(\cdot)$ is computed as: $e(y_{u,i},\hat{y}_{u,i}) = -y_{u,i}\log \hat{y}_{u,i}-(1-y_{u,i})\log \left(1-\hat{y}_{u,i}\right)$. $\frac{1}{1-\hat{y}_p^u}\propto \hat{y}_p^u$ indicates that the weights of negative samples of $\mathcal{U}_\text{portal}'$ increase with users' intent toward the marketing portal. Additionally, $\frac{1}{\hat{y}_b^u}$ reflects decreasing sample confidence as users' block-clicking intent strengthens.

In summary, the final loss of USD is formulated as:
\begin{equation}
    \mathcal{L}_\text{final} = \mathcal{L}_\text{CTR}'+ \alpha * \mathcal{L}_\text{portal} + \beta * \mathcal{L}_\text{block},
\label{eq:loss}
\end{equation}
where $\alpha$ and $\beta$ are hyperparameters that control the weights of the portal and block auxiliary tasks respectively.

\section{Experiments}
\subsection{Experimental Setup}
\subsubsection{Dataset}
% We evaluated USD against baselines using Taobao's industrial dataset capturing real user behaviors.
% which includes two variants of the homepage marketing block: Baiyibutie and Taobaomiaosha. The model was trained on 1.45 billion samples collected over 60 days with 97 million evaluation samples, and served tens of millions of users daily during A/B testing.
We evaluated USD on Taobao's industrial dataset containing 1.45 billion training samples (60-day collection) and 97 million evaluation samples, with A/B testing serving tens of millions of daily users.

\subsubsection{Baseline Models}

We compared with the baseline model \textbf{BASE} (Sec.~\ref{CDM})
and multiple SOTA debiasing methods: classical debiasing model ESMM~\cite{10.1145/3209978.3210104}, counterfactual space modeling method DCMT~\cite{DBLP:conf/icde/ZhuZYL0ZZ0W0W23}, propensity-based approaches ESCM$^2$-IPW \& ESCM$^2$-DR~\cite{10.1145/3477495.3531972}, and pseudo-labeling method NISE~\cite{10.1145/3640457.3688151}.

\subsubsection{Implementation Details}
All models use sampled datasets from Sec.~\ref{subsc:sampling} for fairness. 
Ablation and hyperparameter studies guide online configuration: AdagradDecayV2 optimizer (lr=0.01, batch size=1024), loss weights $\alpha=\beta=0.0001$ (Eq.~\eqref{eq:loss}), debiasing weights (Eq.~\eqref{eq:loss-ctr}) clipped to $[1,15]$.
% \subsubsection{Metrics}
Due to class imbalance in invalid exposure, we use GAUC~\cite{10.1145/3219819.3219823} as primary metric:
$\text{GAUC} = \frac{\sum_{u\in \mathcal{U}}w_u\times \text{AUC}_u}{\sum_{u\in \mathcal{U}}w_u}$, 
where $w_u=1$, exposures, or clicks corresponds to $\text{GAUC}_{\text{avg}}$, $\text{GAUC}_{\text{show}}$, or $\text{GAUC}_{\text{click}}$.
%with $w_u=1$/clicks/exposures defining  $\text{GAUC}_{\text{avg}}$/$\text{GAUC}_{\text{show}}$/$\text{GAUC}_{\text{click}}$.
GAUC better reflects model's real-world online performance through user-group disparities, 
with auxiliary AUC~\cite{FAWCETT2006861} monitoring global performance.
%supplemented by standard AUC~\cite{FAWCETT2006861}.

% GAUC offers a more indicative evaluation through user-group disparities, which we find correlates better with the model’s real-world online performance. We further employ AUC~\cite{FAWCETT2006861}, a standard CTR metric, as our auxiliary offline evaluation metric.

% which we find correlates better with the model's real-world online performance. 
% We further employ AUC~\cite{FAWCETT2006861}, a standard CTR metric, as our auxiliary offline evaluation metric. 

\subsection{Ablation Study}
% \subsubsection{Performance Comparison}
% \label{expres}
% As shown in Table~\ref{tab:expres}, our proposed USD approach outperforms all baselines across all metrics, which validates three key insights: 
% (1) the critical importance of modeling user-specific intents in sampling strategies; (2) multi-intent collaborative modeling facilitates fine-grained dual-debiasing, significantly enhancing model performance on sampled instances; (3) USD demonstrates effectiveness in jointly addressing both SSB and pseudo-exposure bias in large-scale recommendation systems.

% \subsubsection{Ablation Study}
\label{ablationstudy}
Module-wise ablation validates USD's architectural necessity.
\textbf{BASE} (previous online baseline) represents the main model trained on full exposure data without sampling and debiasing. 
%\textbf{BASE} underperforms USD (Fig.~\ref{ablationpic}), confirming sampling and debiasing efficacy.
\textbf{-w/o PS} (click-sampling only) surpasses \textbf{BASE} but underperforms USD, demonstrating that the intent-driven sampler effectively filters invalid exposure samples while preserving confident negatives through latent user intent capture.
Ablated variants (\textbf{-w/o D}: remove the Causal-Based Dual-Debiasing Module (CDM); \textbf{-w/o P}: disable portal-debiasing; \textbf{-w/o B}: disable block-debiasing) exhibit GAUC (highly correlated with online performance) drops, with USD achieving 2.55\%, 0.65\%, and 0.82\% $\text{GAUC}_\text{avg}$ gains, respectively, demonstrating the efficacy of behavioral-derived user intentions for personalized debiasing.

%Component-wise ablation experiments validate the architectural necessity of core modules in USD.
%Module-wise ablation validates USD's structural necessity. 

% \begin{table}[H]
%   \caption{Performance comparison on production datasets.}
%   \label{tab:expresablation}
%   \resizebox{\linewidth}{0.13\linewidth}{
%   \begin{tabular}{c|cccccc}
%     \toprule
%      Models & USD-as & USD-cs & -w/o D & -w/o A & -w/o B & USD\\
%     \midrule
%     GAUC$_\text{avg}$ & 53.63 & 61.19 & 60.20 & 62.10 & 61.93 & \textbf{62.75}\\
%     GAUC$_\text{show}$& 54.96 & 63.40 & 61.93 & 64.19 & 63.73 & \textbf{64.62}\\
%     GAUC$_\text{click}$ & 54.01 & 61.80 & 60.85 & 62.68 & 62.49 & \textbf{63.26}\\
%     AUC & 64.91 & 77.87 & 81.18 & 81.90 & \textbf{83.48} & 82.52\\
%   \bottomrule
% \end{tabular}}
% \end{table}

\begin{table}[H]
  \caption{Performance comparison on production datasets.}
  \label{tab:expres}
  \scalebox{0.85}{
  \begin{tabular}{c|ccccc}
    \toprule
    Method & GAUC$_\text{avg}$ & GAUC$_\text{show}$ & GAUC$_\text{click}$ & AUC\\
    \midrule
    BASE &0.5363  &0.5496  &0.5401  &0.6491 \\
    ESMM & 0.5991 & 0.6166 & 0.6055 & 0.7910\\
    NISE& 0.5436 & 0.5595 & 0.5579 & 0.6996\\
    ESCM$^2$-IPW & 0.5410 & 0.5559 & 0.5553 & 0.6951\\
    ESCM$^2$-DR & 0.5633 & 0.5756 & 0.5721 & 0.7481\\
    DCMT  & 0.6026 & 0.6194 & 0.6080 & 0.8132 \\
    \midrule
    \textbf{USD}  & \textbf{0.6275} & \textbf{0.6462} & \textbf{0.6326} & \textbf{0.8252}\\
  \bottomrule
\end{tabular}
}
\end{table}

\begin{figure}[h]
  \centering
  \includegraphics[width=\linewidth,height = 0.3\linewidth]{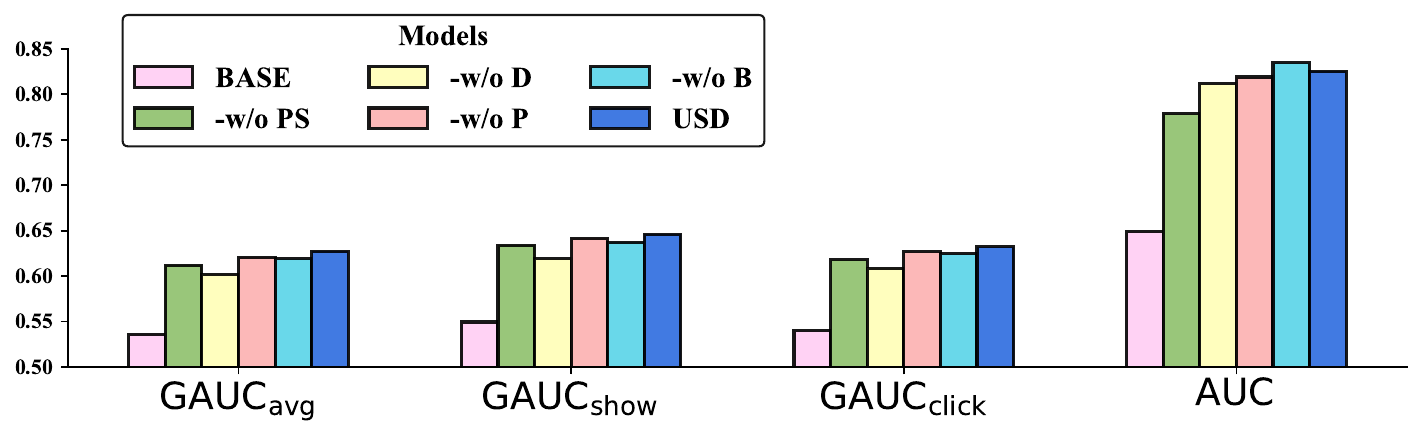}
  \caption{Ablation Study of USD.}
  \Description{Ablation Study of USD.}
  \label{ablationpic}
\end{figure}

\subsection{Online A/B Test}
We ran a week-long online A/B test on Taobao's homepage marketing blocks Baiyibutie and Taobaomiaosha, where USD outperformed the baseline model (\textbf{BASE}, see Sec.~\ref{ablationstudy}) with UCTR increased by 35.4\% and 14.5\% respectively. These substantial gains demonstrate the industrial validity of USD. Furthermore, USD has been fully deployed on Taobao's homepage marketing block.

\section{Conclusions}
We propose the USD framework combining intent-driven sampler and dual-debiasing module to address invalid exposure and click bias on large-scale homepage recommendation.
% We propose the USD framework to address invalid exposure and click bias in large-scale homepage recommendations, integrating an intent-driven sampler to filter invalid exposures and a dual-debiasing module that performs personalized bias correction using user intentions derived from behavioral patterns.
Experiments on industrial datasets and online A/B tests demonstrate its effectiveness. Notably, USD has been fully deployed in Taobao’s homepage marketing blocks, Baiyibutie and Taobaomiaosha.

\section*{PRESENTER BIOGRAPHY}
\textbf{Jiaqi Zheng} is a researcher in the Department of Search and
Recommendation at Taobao \& Tmall Group of Alibaba. He received his master's degree from Beijing University of Posts and Telecommunications. His research focuses on recommendation systems.

\noindent \textbf{Cheng Guo} is a researcher in the Department of Search and Recommendation at Taobao \& Tmall Group of Alibaba. He received his master degree from Tsinghua University, where he was part of THUIR (Information Retrieval Lab at Tsinghua University). His research focuses on information retrieval.

\noindent \textbf{Yi Cao} is the leader of Marketing Algorithm in the Department of Search and Recommendation at Taobao \& Tmall Group of Alibaba. He received his master's degree from Zhejiang University. His research focuses on recommendation systems.

\bibliographystyle{ACM-Reference-Format}
\bibliography{main}

\end{document}